\documentclass[lettersize,journal]{IEEEtran}
\usepackage{amsmath,amsfonts}
\usepackage{algorithmic}
\usepackage{algorithm}
\usepackage{array}
\usepackage[caption=false,font=normalsize,labelfont=sf,textfont=sf]{subfig}
\usepackage{textcomp}
\usepackage{stfloats}
\usepackage{url}
\usepackage{verbatim}
\usepackage{graphicx}
\usepackage{cite}
\usepackage{hyperref}
\usepackage{listings}
\usepackage{xcolor}
\usepackage{pgfplots}
\pgfplotsset{compat=1.18}
\usepackage{booktabs}
\usepackage{multirow}
\definecolor{ieeeblue}{HTML}{0072BD}
\definecolor{ieeeorange}{HTML}{D95319}
\pgfplotsset{
  surveybar/.style={
    width=\columnwidth, bar width=6pt,
    tick label style={font=\scriptsize}, label style={font=\small},
    nodes near coords,
    every node near coord/.append style={font=\scriptsize,
        /pgf/number format/fixed, /pgf/number format/precision=1},
    axis lines*=left, enlarge y limits=0.05,
    xmajorgrids=true, grid style={dotted,gray!40},
  }
}
\begin{document}

\title{
Bridging Artificial Intelligence and Power Systems Education Using a Hands-On Executable Framework}

\author{Junjie Yin,~\IEEEmembership{Graduate Student Member,~IEEE,}
Buxin She,~\IEEEmembership{Member,~IEEE,}\\
Xinyu Feng,~\IEEEmembership{Graduate Student Member,~IEEE,}
Fangxing (Fran) Li,~\IEEEmembership{Fellow,~IEEE}
\thanks{J. Yin, X. Feng, and F. Li are with the EECS Department and the CURENT research center at The University of Tennessee, Knoxville, TN 37996. B. She is with the ECE Department at Kansas State University, Manhattan, KS 66506.}
\thanks{Manuscript received 08/05/2026; revised xx/xx/xxxx.}}

\markboth{Journal of \LaTeX\ Class Files,~Vol.~xx, No.~x, August~2026}%
{Shell \MakeLowercase{\textit{et al.}}: A Sample Article Using IEEEtran.cls for IEEE Journals}


\maketitle
\bstctlcite{IEEEexample:BSTcontrol}

\begin{abstract}
Artificial intelligence (AI) is increasingly central to power and energy systems, supporting modeling, forecasting, optimization, and control. Yet most existing works emphasize specialized applications and offer little reusable material for newcomers or interdisciplinary learners, who increasingly rely on large language models rather than building their own. This gap points to a need for \emph{engineering-grounded AI} (EGAI), in which AI workflows follow established engineering and power-system domain rules rather than acting as task-agnostic black boxes. Motivated by a community survey of researchers and practitioners, which shows $92\%$ report at least one barrier before running an AI model and $94\%$ want a power-specific hands-on course. This paper presents a framework consisting of open, executable module library that lowers the entry barrier for AI in power systems. The modules follow a progressive difficulty ladder that maps core AI concepts onto representative power-system tasks: (i) foundational deep neural network (DNN) templates for function approximation and load-curve fitting; (ii) a domain-coupled convolutional neural network (CNN) power-flow surrogate for a 5-bus system; and (iii) frontier modules on DNN-assisted optimization, deep reinforcement learning (DRL) for battery storage control, and physics-informed neural networks (PINNs) for the swing equation. All modules are released as Jupyter notebooks that run locally or on Google Colab and are delivered through an IEEE online course and IEEE Power \& Energy Society (PES) webinar series. The webinar drew more than $590$ live attendees, which is among the ten most-attended IEEE PES webinars, and over $344$ repository visits within two weeks, reinforcing the survey-based motivation.
\end{abstract}


\section{Introduction}
\subsection{Motivation}
\IEEEPARstart{A}{rtificial} intelligence (AI), as one of the most representative advanced technologies, is now being widely discussed and adopted across many sectors. In the field of power and energy systems, AI has shown strong potential in modeling, forecasting, optimization, and control~\cite{chenReinforcementLearningSelective2022,duchesneRecentDevelopmentsMachine2020}. These capabilities not only accelerate computation but also help address the increasing complexity of modern grids with renewable integration and emerging technologies~\cite{fengGridResilienceExtreme2026}. In daily life, AI has already brought us great convenience, from smart recommendations to voice assistants, which shows how powerful and approachable it can be when proper tools are provided.

However, most of the existing academic work still emphasizes high-level research contributions or specialized applications. While these results are valuable to experts, they are not always accessible to beginners, students, or practitioners from other disciplines. For the general public, getting started with AI is still not straightforward: the workflow may seem abstract, the learning curve is steep, and many open-source codes are either too fragmented or lack context.

Therefore, there is a clear need for simple and generic hands-on codes that can lower the entry barrier. By providing progressive examples which start from basic function approximation, then moving to dataset fitting under noisy conditions, and finally to a real-world inspired power system case, we aim to make AI more approachable. Such examples can serve as stepping stones for students, engineers, and interdisciplinary learners, enabling them to quickly grasp fundamental ideas and adapt the methods to their own problems, whether in power systems, energy forecasting, or other application domains.

\subsection{Related Work}

Recent literature indicates a growing trend of integrating AI into education across multiple stages and disciplines. AI-related curricula and learning activities have been introduced from the secondary school level to higher education, demonstrating the feasibility of early AI exposure as well as advanced AI-supported learning environments \cite{chiuCreationEvaluationPretertiary2022,neumannLLMDrivenChatbotHigher2025}. Within STEM and engineering education, prior studies have explored systematic incorporation of AI fundamentals into curricula and course modules, alongside broader institutional frameworks for adopting generative AI in teaching and learning \cite{haoIntegratingAIEngineering2025,shailendraFrameworkAdoptionGenerative2024}. In parallel, efforts have been made to improve the accessibility and interpretability of AI tools for educators to facilitate classroom deployment \cite{wangMakingAIAccessible2024}. Overall, these works collectively reflect the widespread and increasing integration of AI across educational stages and disciplinary contexts \cite{guedesImpactArtificialIntelligence2025}.




The interplay between AI and power \& energy systems has been extensively surveyed from multiple angles. Existing reviews cover AI-enabled operation, control, and planning in power systems \cite{pandeyApplicationsArtificialIntelligence2023, she2022fusion}, as well as AI techniques across the design, control, and maintenance life-cycle of power-electronics systems \cite{zhaoOverviewArtificialIntelligence2021}. Community perspectives further highlight the mutual coupling between AI workloads and power-electronics infrastructures \cite{chenPowerAIAI2025}, while domain-facing review and vision efforts summarize emerging applications and open challenges within the power-systems community \cite{liReviewVisionAI2022}. In parallel, recent studies advocate open-source ecosystems to improve reproducibility, transparency, and long-term software sustainability \cite{aliPathwayOpenSource2025}, and promote Python-based simulators and digital-twin-like workflows for rapid prototyping and integration with data- and AI-oriented toolchains \cite{haugdalOpenSourcePower2021}.

Despite the growing body of surveys and position papers, many publicly available learning materials remain organized around a single task or narrowly defined use case. Hands-on tutorials often focus on specific problems such as optimal power flow (OPF) \cite{montalvoAIOptimalPower}, while technical exemplars typically emphasize a particular AI or control objective—for example, reinforcement-learning-based adaptive control of converter-interfaced resources under varying grid conditions—rather than a generalizable workflow spanning data handling, model construction, training, and evaluation \cite{fahadDataDrivenAdaptiveControl2025a}. Classical benchmark test systems continue to play an important role in economic and OPF-centered demonstrations \cite{liSmallTestSystems2010}, but such demonstrations do not necessarily translate into reusable, end-to-end coding frameworks that can be readily adapted to a broader range of AI-enabled power-system tasks.

Overall, although existing literature and community resources emphasize reproducible software practices and accessible tooling \cite{aliPathwayOpenSource2025,haugdalOpenSourcePower2021}, few resources provide step-by-step, task-agnostic, hands-on AI programs that beginners can directly reuse and extend. This gap is particularly evident in the context of the rapid expansion of AI methods and their broad applicability across power systems and power electronics, where newcomers often rely primarily on large language models (LLMs) such as ChatGPT, Claude, or Gemini to obtain answers~\cite{yinAIAgentsTaskSimple2026}, rather than developing their own models and experimentation workflows. To address this gap, this work provides a set of educational, hands-on AI modules for function fitting, dataset learning, and power-flow prediction, enabling beginners to actively modify parameters, functions, and higher-level model architectures, and to directly observe their impact on learning outcomes. The proposed modules aim to offer a minimal yet extensible coding baseline that bridges foundational concepts and more advanced AI-enabled power-system analyses.

\subsection{Contribution and Organization}
  


This paper aims to lower the entry barrier for applying AI in power and energy systems through a coherent design framework and a set of hands-on, reusable, and extensible implementations. The main contributions are summarized as follows.

\begin{itemize}
  \item \emph{Needs analysis.} A targeted community survey ($N=52$ valid responses were collected) is conducted to characterize the AI adoption, learning barriers, and resource preferences of power-and-energy researchers and practitioners, and its findings are used to ground the design of the proposed materials.

  \item \emph{Design framework.} A framework is proposed that organizes AI teaching modules along a progressive difficulty ladder and explicitly maps core AI concepts onto representative power-system tasks, all built on a shared, configurable notebook template.

  \item \emph{Executable module library.} Six open, hands-on modules are released across three tiers: foundational deep neural network (DNN) templates for function approximation and load-curve fitting; a domain-coupled convolutional neural network (CNN) power-flow surrogate for a 5-bus system that predicts bus voltages and line active and reactive power flows; and three frontier modules covering DNN-assisted optimization, deep reinforcement learning (DRL) for battery energy storage control, and physics-informed neural networks (PINNs) for the swing equation.

  \item \emph{Reproducibility, integration, and community validation.} All modules are provided as Jupyter notebooks that run locally or on Google Colab and are delivered through an IEEE online course~\cite{Onlinecourse2026} and an IEEE Power \& Energy Society (PES) webinar series~\cite{liWebinarAIPower2026, liWebinarAIPowerslides2026}, enabling learners to move beyond passive reliance on LLMs and to actively explore how parameters, model architectures, and problem formulations affect learning outcomes. Early deployment offers real-world validation: Part~I of the webinar drew more than $590$ live attendees, which ranked the top ten attended IEEE PES webinars, and over $344$ repository visits within two weeks.
\end{itemize}

The rest of this paper is organized as follows.
Section II presents the community survey, educational context, and associated IEEE course and webinar resources.
Section III introduces the proposed hands-on learning framework and module library.
Sections IV--VI describe the released hands-on examples and codes.
Section VII discusses the educational impact, broader applicability, engineering-grounded AI principles, and limitations of the framework.
Section~VIII concludes this paper.

\section{Background and Educational Context}





\subsection{Motivation from a Community Survey on AI Learning Barriers}
\label{sec:survey}

To ground the design of the proposed course in the actual needs of the power and
energy community, we conducted a targeted survey on AI usage, learning barriers,
and resource preferences among academic researchers and industry practitioners.
The instrument comprised nine questions covering respondent background (role,
experience, and technical area), current AI adoption and application domains,
barriers encountered before running an AI model, the perceived relevance of
image-based AI examples, and interest in a hands-on tutorial tailored to power
systems.

\begin{table}[!htbp]
\caption{Respondent Profile and AI Engagement}
\label{tab:profile}\centering
\renewcommand{\arraystretch}{1.15}
\begin{tabular}{@{}l l r r@{}}
\toprule
Attribute & Category & $n$ & \% \\
\midrule
\multirow{4}{*}{Primary role (Q1)}
 & Graduate student             & 20 & 38.5 \\
 & Faculty / research scientist & 14 & 26.9 \\
 & Industry engineer            & 10 & 19.2 \\
 & Undergraduate / other        &  8 & 15.4 \\
\midrule
\multirow{4}{*}{Experience (Q2)}
 & $<3$ years   & 15 & 28.8 \\
 & 3--5 years   &  8 & 15.4 \\
 & 5--10 years  & 15 & 28.8 \\
 & $>10$ years  & 14 & 26.9 \\
\midrule
\multirow{4}{*}{\shortstack[l]{AI projects in\\past 2\,yr (Q4)}}
 & More than 2 projects & 15 & 28.8 \\
 & 2 projects           &  9 & 17.3 \\
 & 1 project            & 15 & 28.8 \\
 & Never used           & 13 & 25.0 \\
\midrule
\multirow{3}{*}{\shortstack[l]{Want one-click\\hands-on course (Q8)}}
 & Yes   & 34 & 65.4 \\
 & Maybe & 15 & 28.8 \\
 & No    &  3 &  5.8 \\
\bottomrule
\end{tabular}
\end{table}

\subsubsection{Broad, cross-domain interest in AI}
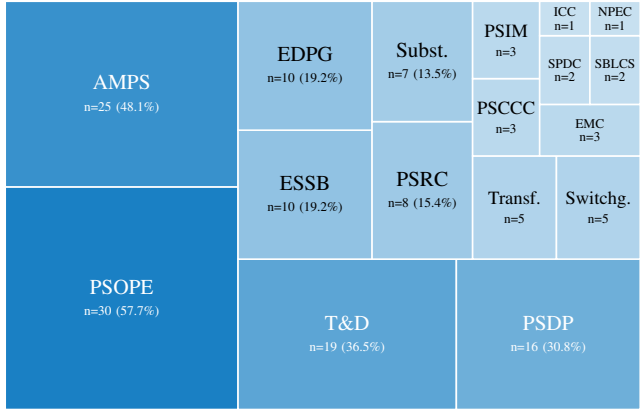
\begin{figure}[!htbp]\centering
\begin{tikzpicture}[x=1cm,y=1cm]
  \fill[ieeeblue!90] (0,0) rectangle (3.08,2.945);
  \draw[white,line width=0.6pt] (0,0) rectangle (3.08,2.945);
  \node[align=center,text=white,font=\footnotesize] at (1.54,1.473) {PSOPE\\[-1pt]{\tiny n{=}30 (57.7\%)}};
  \fill[ieeeblue!78] (0,2.945) rectangle (3.08,5.4);
  \draw[white,line width=0.6pt] (0,2.945) rectangle (3.08,5.4);
  \node[align=center,text=white,font=\footnotesize] at (1.54,4.173) {AMPS\\[-1pt]{\tiny n{=}25 (48.1\%)}};
  \fill[ieeeblue!64] (3.08,0) rectangle (5.968,1.989);
  \draw[white,line width=0.6pt] (3.08,0) rectangle (5.968,1.989);
  \node[align=center,text=white,font=\footnotesize] at (4.524,0.995) {T\&D\\[-1pt]{\tiny n{=}19 (36.5\%)}};
  \fill[ieeeblue!57] (5.968,0) rectangle (8.4,1.989);
  \draw[white,line width=0.6pt] (5.968,0) rectangle (8.4,1.989);
  \node[align=center,text=white,font=\footnotesize] at (7.184,0.995) {PSDP\\[-1pt]{\tiny n{=}16 (30.8\%)}};
  \fill[ieeeblue!43] (3.08,1.989) rectangle (4.853,3.695);
  \draw[white,line width=0.6pt] (3.08,1.989) rectangle (4.853,3.695);
  \node[align=center,text=black,font=\footnotesize] at (3.967,2.842) {ESSB\\[-1pt]{\tiny n{=}10 (19.2\%)}};
  \fill[ieeeblue!43] (3.08,3.695) rectangle (4.853,5.4);
  \draw[white,line width=0.6pt] (3.08,3.695) rectangle (4.853,5.4);
  \node[align=center,text=black,font=\footnotesize] at (3.967,4.547) {EDPG\\[-1pt]{\tiny n{=}10 (19.2\%)}};
  \fill[ieeeblue!38] (4.853,1.989) rectangle (6.183,3.808);
  \draw[white,line width=0.6pt] (4.853,1.989) rectangle (6.183,3.808);
  \node[align=center,text=black,font=\footnotesize] at (5.518,2.899) {PSRC\\[-1pt]{\tiny n{=}8 (15.4\%)}};
  \fill[ieeeblue!36] (4.853,3.808) rectangle (6.183,5.4);
  \draw[white,line width=0.6pt] (4.853,3.808) rectangle (6.183,5.4);
  \node[align=center,text=black,font=\footnotesize] at (5.518,4.604) {Subst.\\[-1pt]{\tiny n{=}7 (13.5\%)}};
  \fill[ieeeblue!31] (6.183,1.989) rectangle (7.292,3.354);
  \draw[white,line width=0.6pt] (6.183,1.989) rectangle (7.292,3.354);
  \node[align=center,text=black,font=\scriptsize] at (6.738,2.672) {Transf.\\[-1pt]{\tiny n{=}5}};
  \fill[ieeeblue!31] (7.292,1.989) rectangle (8.4,3.354);
  \draw[white,line width=0.6pt] (7.292,1.989) rectangle (8.4,3.354);
  \node[align=center,text=black,font=\scriptsize] at (7.846,2.672) {Switchg.\\[-1pt]{\tiny n{=}5}};
  \fill[ieeeblue!27] (6.183,3.354) rectangle (7.07,4.377);
  \draw[white,line width=0.6pt] (6.183,3.354) rectangle (7.07,4.377);
  \node[align=center,text=black,font=\scriptsize] at (6.627,3.865) {PSCCC\\[-1pt]{\tiny n{=}3}};
  \fill[ieeeblue!27] (6.183,4.377) rectangle (7.07,5.4);
  \draw[white,line width=0.6pt] (6.183,4.377) rectangle (7.07,5.4);
  \node[align=center,text=black,font=\scriptsize] at (6.627,4.888) {PSIM\\[-1pt]{\tiny n{=}3}};
  \fill[ieeeblue!27] (7.07,3.354) rectangle (8.4,4.036);
  \draw[white,line width=0.6pt] (7.07,3.354) rectangle (8.4,4.036);
  \node[align=center,text=black,font=\tiny] at (7.735,3.695) {EMC\\[-1pt]{\tiny n{=}3}};
  \fill[ieeeblue!24] (7.07,4.036) rectangle (7.735,4.945);
  \draw[white,line width=0.6pt] (7.07,4.036) rectangle (7.735,4.945);
  \node[align=center,text=black,font=\tiny] at (7.402,4.491) {SPDC\\[-1pt]{\tiny n{=}2}};
  \fill[ieeeblue!24] (7.735,4.036) rectangle (8.4,4.945);
  \draw[white,line width=0.6pt] (7.735,4.036) rectangle (8.4,4.945);
  \node[align=center,text=black,font=\tiny] at (8.067,4.491) {SBLCS\\[-1pt]{\tiny n{=}2}};
  \fill[ieeeblue!22] (7.07,4.945) rectangle (7.735,5.4);
  \draw[white,line width=0.6pt] (7.07,4.945) rectangle (7.735,5.4);
  \node[align=center,text=black,font=\tiny] at (7.403,5.173) {ICC\\[-1pt]{\tiny n{=}1}};
  \fill[ieeeblue!22] (7.735,4.945) rectangle (8.4,5.4);
  \draw[white,line width=0.6pt] (7.735,4.945) rectangle (8.4,5.4);
  \node[align=center,text=black,font=\tiny] at (8.068,5.173) {NPEC\\[-1pt]{\tiny n{=}1}};
\end{tikzpicture}
\caption{Treemap of respondents' research and work areas across the IEEE PES
technical committees (Q3, multiple selections).}
\label{fig:areas}
\end{figure}

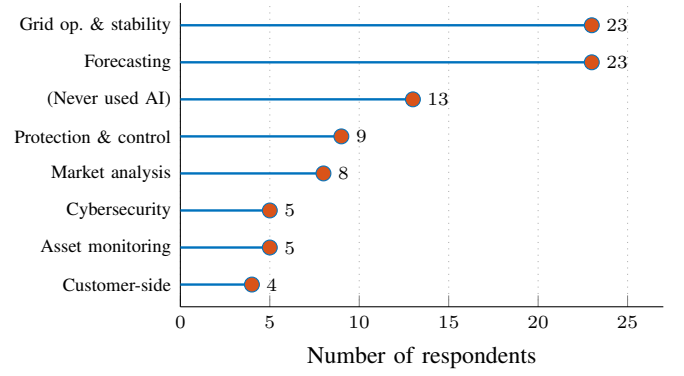
\begin{figure}[!htbp]\centering
\begin{tikzpicture}
\begin{axis}[
  width=0.9\columnwidth, height=5.6cm,
  xmin=0, xmax=27, ymin=0.4, ymax=8.6,
  xtick={0,5,10,15,20,25},
  ytick={1,2,3,4,5,6,7,8},
  yticklabels={{Customer-side},{Asset monitoring},{Cybersecurity},%
               {Market analysis},{Protection \& control},{(Never used AI)},%
               {Forecasting},{Grid op.\ \& stability}},
  xlabel={Number of respondents},
  xmajorgrids=true, grid style={dotted,black!30},
  tick label style={font=\scriptsize}, label style={font=\small},
  axis lines*=left, clip=false,
]
\addplot[xcomb, draw=ieeeblue, line width=0.9pt, mark=none] coordinates {
  (4,1) (5,2) (5,3) (8,4) (9,5) (13,6) (23,7) (23,8)};
\addplot[only marks, mark=*, mark size=2.8pt, draw=ieeeblue, fill=ieeeorange,
  nodes near coords, point meta=x,
  every node near coord/.append style={anchor=west, xshift=2pt, font=\scriptsize,
     /pgf/number format/.cd, fixed, precision=0}] coordinates {
  (4,1) (5,2) (5,3) (8,4) (9,5) (13,6) (23,7) (23,8)};
\end{axis}
\end{tikzpicture}
\caption{Power-system application domains in which respondents have applied AI
(Q5, multiple selections).}
\label{fig:apps}
\end{figure}

Table~\ref{tab:profile} summarizes the respondent profile, where Q\# denotes the question index in the survey and $n$ denotes the number of responses in each category. The sample spans
graduate students ($38.5\%$), faculty and research scientists ($26.9\%$),
industry engineers ($19.2\%$), and undergraduates, with experience distributed
across all ranges from under three years to more than ten years. As shown in
Fig.~\ref{fig:areas}, respondents' research and work areas cover essentially all
IEEE PES technical committees, led by power system
operation, planning and economics ($57.7\%$), analytic methods ($48.1\%$),
transmission and distribution ($36.5\%$), and dynamic performance ($30.8\%$).
This breadth indicates that interest in AI is not confined to a single
specialization but is shared across the discipline. Consistently,
Fig.~\ref{fig:apps} shows that respondents have already applied AI to a wide
range of tasks: most frequently forecasting (23 counts) and grid operation and stability (23 counts), followed by protection and control, market analysis, asset
monitoring, and cybersecurity. But there is still 13 counts responses selected ``Not used AI", which motivated us to provide more entry-level friendly tutotials to them.

\subsubsection{Persistent barriers, even for experienced users}
\begin{figure}[!htbp]\centering
\begin{tikzpicture}[x=1cm,y=1cm]
  \node[anchor=south,align=center,font=\scriptsize] at (2.8,0.07) {Graduate\\[-1pt]{\tiny n{=}20}};
  \node[anchor=south,align=center,font=\scriptsize] at (4.1,0.07) {Faculty\\[-1pt]{\tiny n{=}14}};
  \node[anchor=south,align=center,font=\scriptsize] at (5.4,0.07) {Industry\\[-1pt]{\tiny n{=}10}};
  \node[anchor=south,align=center,font=\scriptsize] at (6.7,0.07) {Undergrad\\[-1pt]{\tiny n{=}8}};
  \node[anchor=south,align=center,font=\scriptsize] at (8,0.07) {Overall\\[-1pt]{\tiny N{=}52}};
  \node[anchor=east,font=\scriptsize] at (2.07,-0.36) {Lack of background};
  \fill[black!9,rounded corners=0.5pt] (2.25,-0.49) rectangle (3.09,-0.23);
  \fill[ieeeblue!82,rounded corners=0.5pt] (2.25,-0.49) rectangle (2.628,-0.23);
  \node[anchor=east,font=\scriptsize] at (3.4,-0.36) {9};
  \fill[black!9,rounded corners=0.5pt] (3.55,-0.49) rectangle (4.39,-0.23);
  \fill[ieeeblue!82,rounded corners=0.5pt] (3.55,-0.49) rectangle (3.91,-0.23);
  \node[anchor=east,font=\scriptsize] at (4.7,-0.36) {6};
  \fill[black!9,rounded corners=0.5pt] (4.85,-0.49) rectangle (5.69,-0.23);
  \fill[ieeeblue!82,rounded corners=0.5pt] (4.85,-0.49) rectangle (5.27,-0.23);
  \node[anchor=east,font=\scriptsize] at (6,-0.36) {5};
  \fill[black!9,rounded corners=0.5pt] (6.15,-0.49) rectangle (6.99,-0.23);
  \fill[ieeeblue!82,rounded corners=0.5pt] (6.15,-0.49) rectangle (6.675,-0.23);
  \node[anchor=east,font=\scriptsize] at (7.3,-0.36) {5};
  \fill[black!9,rounded corners=0.5pt] (7.45,-0.49) rectangle (8.29,-0.23);
  \fill[ieeeorange!85,rounded corners=0.5pt] (7.45,-0.49) rectangle (7.854,-0.23);
  \node[anchor=east,font=\scriptsize\bfseries] at (8.6,-0.36) {25};
  \node[anchor=east,font=\scriptsize] at (2.07,-1.08) {Hardware limits};
  \fill[black!9,rounded corners=0.5pt] (2.25,-1.21) rectangle (3.09,-0.95);
  \fill[ieeeblue!82,rounded corners=0.5pt] (2.25,-1.21) rectangle (2.502,-0.95);
  \node[anchor=east,font=\scriptsize] at (3.4,-1.08) {6};
  \fill[black!9,rounded corners=0.5pt] (3.55,-1.21) rectangle (4.39,-0.95);
  \fill[ieeeblue!82,rounded corners=0.5pt] (3.55,-1.21) rectangle (4.03,-0.95);
  \node[anchor=east,font=\scriptsize] at (4.7,-1.08) {8};
  \fill[black!9,rounded corners=0.5pt] (4.85,-1.21) rectangle (5.69,-0.95);
  \fill[ieeeblue!82,rounded corners=0.5pt] (4.85,-1.21) rectangle (5.186,-0.95);
  \node[anchor=east,font=\scriptsize] at (6,-1.08) {4};
  \fill[black!9,rounded corners=0.5pt] (6.15,-1.21) rectangle (6.99,-0.95);
  \fill[ieeeblue!82,rounded corners=0.5pt] (6.15,-1.21) rectangle (6.465,-0.95);
  \node[anchor=east,font=\scriptsize] at (7.3,-1.08) {3};
  \fill[black!9,rounded corners=0.5pt] (7.45,-1.21) rectangle (8.29,-0.95);
  \fill[ieeeorange!85,rounded corners=0.5pt] (7.45,-1.21) rectangle (7.789,-0.95);
  \node[anchor=east,font=\scriptsize\bfseries] at (8.6,-1.08) {21};
  \node[anchor=east,font=\scriptsize] at (2.07,-1.8) {Env.\ setup};
  \fill[black!9,rounded corners=0.5pt] (2.25,-1.93) rectangle (3.09,-1.67);
  \fill[ieeeblue!82,rounded corners=0.5pt] (2.25,-1.93) rectangle (2.502,-1.67);
  \node[anchor=east,font=\scriptsize] at (3.4,-1.8) {6};
  \fill[black!9,rounded corners=0.5pt] (3.55,-1.93) rectangle (4.39,-1.67);
  \fill[ieeeblue!82,rounded corners=0.5pt] (3.55,-1.93) rectangle (3.85,-1.67);
  \node[anchor=east,font=\scriptsize] at (4.7,-1.8) {5};
  \fill[black!9,rounded corners=0.5pt] (4.85,-1.93) rectangle (5.69,-1.67);
  \fill[ieeeblue!82,rounded corners=0.5pt] (4.85,-1.93) rectangle (5.018,-1.67);
  \node[anchor=east,font=\scriptsize] at (6,-1.8) {2};
  \fill[black!9,rounded corners=0.5pt] (6.15,-1.93) rectangle (6.99,-1.67);
  \fill[ieeeblue!82,rounded corners=0.5pt] (6.15,-1.93) rectangle (6.36,-1.67);
  \node[anchor=east,font=\scriptsize] at (7.3,-1.8) {2};
  \fill[black!9,rounded corners=0.5pt] (7.45,-1.93) rectangle (8.29,-1.67);
  \fill[ieeeorange!85,rounded corners=0.5pt] (7.45,-1.93) rectangle (7.692,-1.67);
  \node[anchor=east,font=\scriptsize\bfseries] at (8.6,-1.8) {15};
  \node[anchor=east,font=\scriptsize] at (2.07,-2.52) {OS differences};
  \fill[black!9,rounded corners=0.5pt] (2.25,-2.65) rectangle (3.09,-2.39);
  \fill[ieeeblue!82,rounded corners=0.5pt] (2.25,-2.65) rectangle (2.376,-2.39);
  \node[anchor=east,font=\scriptsize] at (3.4,-2.52) {3};
  \fill[black!9,rounded corners=0.5pt] (3.55,-2.65) rectangle (4.39,-2.39);
  \fill[ieeeblue!82,rounded corners=0.5pt] (3.55,-2.65) rectangle (3.67,-2.39);
  \node[anchor=east,font=\scriptsize] at (4.7,-2.52) {2};
  \fill[black!9,rounded corners=0.5pt] (4.85,-2.65) rectangle (5.69,-2.39);
  \fill[ieeeblue!82,rounded corners=0.5pt] (4.85,-2.65) rectangle (5.018,-2.39);
  \node[anchor=east,font=\scriptsize] at (6,-2.52) {2};
  \fill[black!9,rounded corners=0.5pt] (6.15,-2.65) rectangle (6.99,-2.39);
  \fill[ieeeblue!82,rounded corners=0.5pt] (6.15,-2.65) rectangle (6.255,-2.39);
  \node[anchor=east,font=\scriptsize] at (7.3,-2.52) {1};
  \fill[black!9,rounded corners=0.5pt] (7.45,-2.65) rectangle (8.29,-2.39);
  \fill[ieeeorange!85,rounded corners=0.5pt] (7.45,-2.65) rectangle (7.579,-2.39);
  \node[anchor=east,font=\scriptsize\bfseries] at (8.6,-2.52) {8};
  \node[anchor=east,font=\scriptsize] at (2.07,-3.24) {Debugging};
  \fill[black!9,rounded corners=0.5pt] (2.25,-3.37) rectangle (3.09,-3.11);
  \fill[ieeeblue!82,rounded corners=0.5pt] (2.25,-3.37) rectangle (2.376,-3.11);
  \node[anchor=east,font=\scriptsize] at (3.4,-3.24) {3};
  \fill[black!9,rounded corners=0.5pt] (3.55,-3.37) rectangle (4.39,-3.11);
  \node[anchor=east,font=\scriptsize] at (4.7,-3.24) {0};
  \fill[black!9,rounded corners=0.5pt] (4.85,-3.37) rectangle (5.69,-3.11);
  \node[anchor=east,font=\scriptsize] at (6,-3.24) {0};
  \fill[black!9,rounded corners=0.5pt] (6.15,-3.37) rectangle (6.99,-3.11);
  \fill[ieeeblue!82,rounded corners=0.5pt] (6.15,-3.37) rectangle (6.255,-3.11);
  \node[anchor=east,font=\scriptsize] at (7.3,-3.24) {1};
  \fill[black!9,rounded corners=0.5pt] (7.45,-3.37) rectangle (8.29,-3.11);
  \fill[ieeeorange!85,rounded corners=0.5pt] (7.45,-3.37) rectangle (7.515,-3.11);
  \node[anchor=east,font=\scriptsize\bfseries] at (8.6,-3.24) {4};
  \draw[black!45,line width=0.6pt] (2.15,0) -- (8.65,0);
  \draw[black!45,line width=0.6pt] (2.15,-3.6) -- (8.65,-3.6);
  \draw[black!12,line width=0.4pt] (2.15,-0.72) -- (8.65,-0.72);
  \draw[black!12,line width=0.4pt] (2.15,-1.44) -- (8.65,-1.44);
  \draw[black!12,line width=0.4pt] (2.15,-2.16) -- (8.65,-2.16);
  \draw[black!12,line width=0.4pt] (2.15,-2.88) -- (8.65,-2.88);
  \draw[black!45,line width=0.6pt] (7.35,0) -- (7.35,-3.6);
\end{tikzpicture}
\caption{Barriers reported before running an AI model, broken down by respondent
role (Q6, multiple selections).}
\label{fig:barriers}
\end{figure}
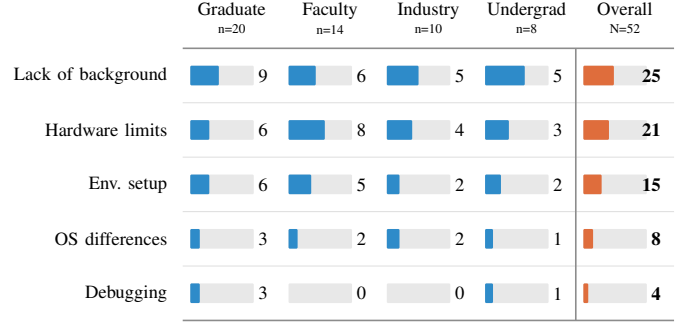

Despite this broad interest, adoption is uneven and accompanied by persistent
barriers. Three quarters of respondents ($75\%$) had used AI in at least one
project in the past two years, while one quarter had never applied AI beyond
conversational tools (Table~\ref{tab:profile}). Crucially, these barriers are
not limited to newcomers. As shown in Fig.~\ref{fig:barriers}, the most
frequently reported obstacles are insufficient background knowledge and not
knowing where to start ($48.1\%$), hardware
limitations such as GPU and memory constraints ($40.4\%$), and
environment-configuration issues including framework installation and Python
version conflicts ($28.8\%$). Each bar in Fig.~\ref{fig:barriers} reports
the within-group share, the number of respondents in a role group selecting a
barrier divided by that group's size, so that roles of unequal size remain
comparable, whereas the ``Overall" column is computed over all 
respondents rather than by averaging the group shares. Insufficient background
knowledge is the leading barrier for every role, while hardware limitations are
most acute for faculty and research scientists ($57\%$). In
aggregate, $92\%$ of respondents reported at least one barrier, and among the $39$ respondents who had already completed one or more AI
projects, $95\%$ still reported a barrier. These findings highlight the
importance of execution-ready learning materials that minimize local setup
requirements.

\subsubsection{Limited relevance of generic, image-based examples}
\begin{figure}[!htbp]\centering
\begin{tikzpicture}[x=1cm,y=1cm]
  \fill[ieeeblue!90] (90:1) -- (90:1.85) arc(90:48.559:1.85) -- (48.559:1) arc(48.559:90:1) -- cycle;
  \node[font=\scriptsize,text=white] at (69.279:1.425) {11.5\%};
  \fill[ieeeblue!68] (48.559:1) -- (48.559:1.85) arc(48.559:-76.126:1.85) -- (-76.126:1) arc(-76.126:48.559:1) -- cycle;
  \node[font=\scriptsize,text=white] at (-13.784:1.425) {34.6\%};
  \fill[ieeeblue!48] (-76.126:1) -- (-76.126:1.85) arc(-76.126:-179.91:1.85) -- (-179.91:1) arc(-179.91:-76.126:1) -- cycle;
  \node[font=\scriptsize,text=white] at (-128.018:1.425) {28.8\%};
  \fill[ieeeblue!30] (-179.91:1) -- (-179.91:1.85) arc(-179.91:-235.405:1.85) -- (-235.405:1) arc(-235.405:-179.91:1) -- cycle;
  \node[font=\scriptsize,text=black] at (-207.658:1.425) {15.4\%};
  \fill[ieeeblue!16] (-235.405:1) -- (-235.405:1.85) arc(-235.405:-270:1.85) -- (-270:1) arc(-270:-235.405:1) -- cycle;
  \node[font=\scriptsize,text=black] at (-252.703:1.425) {9.6\%};
  \fill[ieeeblue!90] (2.4,1) rectangle (2.72,1.32);
  \draw[black!40,line width=0.3pt] (2.4,1) rectangle (2.72,1.32);
  \node[anchor=west,font=\scriptsize] at (2.82,1.16) {Very relevant};
  \fill[ieeeblue!68] (2.4,0.48) rectangle (2.72,0.8);
  \draw[black!40,line width=0.3pt] (2.4,0.48) rectangle (2.72,0.8);
  \node[anchor=west,font=\scriptsize] at (2.82,0.64) {Somewhat relevant};
  \fill[ieeeblue!48] (2.4,-0.04) rectangle (2.72,0.28);
  \draw[black!40,line width=0.3pt] (2.4,-0.04) rectangle (2.72,0.28);
  \node[anchor=west,font=\scriptsize] at (2.82,0.12) {Neutral};
  \fill[ieeeblue!30] (2.4,-0.56) rectangle (2.72,-0.24);
  \draw[black!40,line width=0.3pt] (2.4,-0.56) rectangle (2.72,-0.24);
  \node[anchor=west,font=\scriptsize] at (2.82,-0.4) {Slightly irrelevant};
  \fill[ieeeblue!16] (2.4,-1.08) rectangle (2.72,-0.76);
  \draw[black!40,line width=0.3pt] (2.4,-1.08) rectangle (2.72,-0.76);
  \node[anchor=west,font=\scriptsize] at (2.82,-0.92) {Not relevant at all};
\end{tikzpicture}
\caption{Perceived relevance of image-based AI examples (e.g., MNIST
handwritten-digit recognition) to power-system applications (Q7, single choice).}
\label{fig:mnist}
\end{figure}
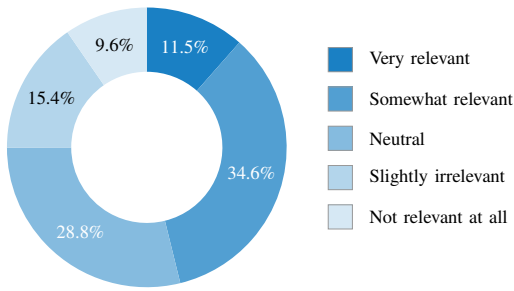

A second motivating finding concerns the limited relevance of generic AI
tutorials to power-system problems. As shown in Fig.~\ref{fig:mnist}, when asked
how relevant image-based examples such as MNIST handwritten-digit recognition
are to AI applications in power systems, only $11.5\%$ of respondents rated them
``very relevant,'' whereas $53.8\%$ were neutral-to-negative. This gap echoes a
recurring theme in the open-ended responses: publicly available tutorials are
perceived as over-focused on computer-vision or text tasks, while power-system
problems are dominated by time series, waveforms, power-flow quantities, and
dynamic responses. Respondents repeatedly requested domain-specific, hands-on
examples grounded in realistic power-system data, together with access to
realistic datasets, affordable or cloud-based computation, and evidence of
reliability.

\subsubsection{Clear demand for a tailored, low-friction learning path}
These needs translate into a clear demand for a tailored learning path. When
asked whether they would use a one-click, hands-on AI tutorial designed
specifically for power and energy systems, $65.4\%$ answered ``yes'' and a
further $28.8\%$ answered ``maybe,'' for a combined $94.2\%$
(Table~\ref{tab:profile}). Together with the qualitative feedback, these results
directly motivate the hands-on, execution-ready course developed in this paper:
by combining cloud-executable Jupyter notebooks, modular and physically
interpretable examples, and minimal local setup, the proposed course targets the
most frequently reported barriers and the demonstrated demand for
power-system-specific AI education.

\subsection{Overview of the Online Course and Webinar}
\subsubsection{IEEE Online Course}
An example of recent educational efforts is the online course \emph{AI for Power and Energy Systems: Applications, Challenges, and Opportunities}~\cite{Onlinecourse2026}, hosted by IEEE. The course introduces AI techniques for accelerating modeling and control tasks in power and energy systems and is accompanied by example implementations provided as Jupyter notebooks. Representative examples include DNN training for function approximation and CNN models applied to a small power system, demonstrating how AI models can be connected to power-system problems in a computational environment.

\subsubsection{IEEE PES Webinar Series}
The same hands-on modules are delivered through the IEEE PES webinar \emph{AI for Power System Applications: Hands-on Examples, Selected Applications, and Perspectives}~\cite{liWebinarAIPower2026, liWebinarAIPowerslides2026}. Part~I presents the foundational and domain-coupled modules (01--03), while the frontier modules (04--06) will follow in Part~II. 
This early engagement provides direct, real-world feedback that complements the community survey and corroborates the demand for reproducible, power-system-specific AI learning materials.

\subsection{Generic Codes as Hands-on Learning Tools}


Hands-on practice plays a critical role in reinforcing theoretical concepts and developing practical skills in AI and data-driven modeling. While domain-specific examples are useful for illustrating particular applications, generic and reusable code templates can further enhance learning by allowing users to focus on core AI principles rather than problem-specific details. Such generic codes can be readily adapted beyond power systems to other engineering and applied science domains with similar data-driven characteristics.

Well-structured hands-on examples can also accommodate learners at different levels. For beginners, simplified and modular code provides an accessible entry point to AI programming, enabling experimentation with parameters, functions, and model architectures. For more advanced learners, the same codes can serve as a baseline for extending models, incorporating additional data, or exploring more sophisticated learning frameworks. As a result, generic hands-on codes can function as a common starting point for AI education across disciplines, bridging foundational learning and advanced applications.

\section{A Framework for Hands-On AI Modules in Power Systems}

The survey in Section~\ref{sec:survey} indicates that learning barriers are nearly universal and that generic, image-based examples are perceived as weakly relevant to power engineering. These findings motivate an approach that is more than a loose collection of scripts: a \emph{framework} in which every module deliberately couples a core AI concept to a representative power-system task, so that learners always see \emph{why} a method matters. This section describes the two knowledge maps that underpin the framework (Section~\ref{subsec:maps}), the design philosophy that turns them into reusable modules (Section~\ref{subsec:philosophy}), and the resulting module library at a glance (Section~\ref{subsec:library}) \cite{yinIEEECourseAIPower2025}.

\begin{table*}[!htbp]
\caption{Overview of the hands-on module library (file names in parentheses)~\cite{yinIEEECourseAIPower2025}.}
\label{tab:modules}
\centering
\footnotesize
\setlength{\tabcolsep}{4pt}
\renewcommand{\arraystretch}{1.3}
\begin{tabular}{@{}>{\raggedright\arraybackslash}p{0.68in} >{\raggedright\arraybackslash}p{1.85in} >{\raggedright\arraybackslash}p{1.35in} >{\raggedright\arraybackslash}p{1.38in} >{\raggedright\arraybackslash}p{1.12in}@{}}
\toprule
Tier & Module (files) & Core AI concept & Power-system task & System / data \\
\midrule
\multirow{2}{*}[-21pt]{\shortstack[l]{Tier~1\\(Foundational)}} & 01 DNN function fitting \newline (\texttt{DNN\_sin.ipynb} \newline \texttt{DNN\_template.ipynb}) & Nonlinear regression (MLP) & Function approximation & Analytic $y=f(x)$ \\
\cmidrule(lr){2-5}
 & 02 DNN load curve fitting \newline (\texttt{DNN\_load\_curve.ipynb} \newline \texttt{1min\_load\_linear.xlsx}) & Data-driven regression & Load-curve prediction & 1-min load data \\
\midrule
Tier~2 \newline (Domain-coupled) & 03 CNN for 5 bus system \newline (\texttt{CNN\_for\_5\_bus\_system.ipynb}) & 1-D convolution, multi-output & Power-flow surrogate & PJM 5-bus, Monte-Carlo power flow \\
\midrule
\multirow{3}{*}[-21pt]{\shortstack[l]{Tier~3\\(Frontier)}} & 04 DNN Assisted Optimization \newline (\texttt{DNN\_Assisted\_Opti.ipynb}) & Learned constraint via MILP & Security-constrained decisions & Synthetic feasibility set \\
\cmidrule(lr){2-5}
 & 05 DRL BESS \newline (\texttt{DRL\_BESS.ipynb}) & Value-based control (DQN) & Battery scheduling & 24-h load, ToU price \\
\cmidrule(lr){2-5}
 & 06 PINN Frequency \newline (\texttt{PINN\_Frequency.ipynb}) & Physics-informed training & Frequency dynamics & Linearized swing equation \\
\bottomrule
\end{tabular}
\end{table*}

\subsection{Coupling an AI Knowledge Map to a Power-System Knowledge Map}
\label{subsec:maps}

The framework is organized around two complementary knowledge maps. The \emph{AI knowledge map} orders methods by the type of problem they solve: nonlinear regression and function approximation with feed-forward DNNs; spatial or structured feature extraction with CNNs; constrained decision making, where a learned model is embedded inside a mathematical program; sequential decision making under uncertainty with DRL; and physics-constrained learning with PINNs. The \emph{power-system knowledge map} orders tasks by function: load and renewable forecasting; power-flow analysis and state estimation; security-constrained dispatch and scheduling; storage and demand-side control; and electromechanical dynamics and frequency response.

The core idea of the framework is to \emph{pair} entries across the two maps rather than teach them in isolation. Function approximation is paired with load-curve prediction; convolution is paired with a power-flow surrogate; a learned constraint is paired with security-constrained decision making; DRL is paired with battery storage control; and physics-informed training is paired with the swing equation. This pairing directly addresses the survey finding that image-classification exemplars feel disconnected from power engineering: each module keeps the same generic AI workflow but anchors it in a recognizable power-system problem.

\subsection{Design Philosophy}
\label{subsec:philosophy}

Four principles turn the two maps into a coherent set of modules.

\emph{Progressive difficulty.} The modules are arranged in three tiers of increasing conceptual and engineering complexity: a \emph{foundational} tier that establishes the supervised-learning workflow, a \emph{domain-coupled} tier that connects a model to a physics-based simulator, and a \emph{frontier} tier that reaches methods at the current research edge.

\emph{Unified template.} Every module follows the same notebook skeleton: a \texttt{Settings} block (Section~0), data generation, model construction, training, evaluation, and visualization. Therefore, a learner who understands one module can navigate any other by editing a single configuration block.

\emph{Minimal yet extensible.} Each notebook ships with small defaults that run in seconds on a laptop or in Colab, together with explicit notes on how to scale up (e.g., more samples, larger networks, or bigger test systems). This keeps the first run fast while leaving a clear path to realistic problem sizes.

\emph{Reproducibility and active exploration.} All modules are released under a permissive license with a pinned environment specification and one-click Colab execution, so that results can be reproduced exactly. By exposing parameters, architectures, and problem formulations, the modules encourage learners to experiment actively rather than to rely passively on LLMs for answers.

\subsection{The Module Library at a Glance}
\label{subsec:library}

Table~\ref{tab:modules} summarizes the six modules, their AI concepts, the power-system tasks they target, and the underlying systems or data. The remainder of the paper presents the library tier by tier: Section~\ref{sec:tier1} details the foundational DNN templates as a worked exemplar; Section~\ref{sec:tier2} details the domain-coupled CNN power-flow surrogate as a second exemplar; and Section~\ref{sec:tier3} presents the three frontier modules more concisely, emphasizing how each extends the shared workflow to a new class of power-system problem. Full source code for every module is available in the accompanying open repository~\cite{yinIEEECourseAIPower2025}.

\section{Foundational Modules (Tier~1): Generic DNN Templates}
\label{sec:tier1}

The foundational tier centers on a single feed-forward DNN template, implemented in \texttt{DNN\_template.ipynb}, that serves as a generic starting point for one-dimensional regression. The same template is exercised in two modes that share an identical pipeline and differ only in the data source: \emph{analytic function approximation}, where samples come from a user-selected closed-form function, and \emph{dataset / load-curve fitting} (\texttt{DNN\_load\_curve}, Section~\ref{subsec:loadcurve}), where they come from measured data. The design logic is that moving from a textbook function to real data changes only the data source, not the AI workflow; accordingly, this section emphasizes that structure and leaves the step-by-step usage to the notebooks.

\subsection{Problem Formulation}

The module approximates an unknown nonlinear mapping $y=f(x)$ with scalar $x,y\in\mathbb{R}$, a formulation that covers a broad class of one-dimensional regression problems. To illustrate generality, the \emph{Settings} block (Section~0) exposes several selectable function modes: \texttt{sin}, \texttt{cos}, \texttt{exp}, \texttt{log}, \texttt{poly3} (a cubic polynomial), and a user-defined \texttt{custom} mode. With this design, the learning task can be changed without touching the training pipeline.

\subsection{Network Architecture and Training Setup}

A fully connected DNN, i.e., a multilayer perceptron (MLP), maps the scalar input $x$ to a scalar prediction $\hat{y}$, implemented in TensorFlow/Keras with a mean-squared-error (MSE) loss and the Adam optimizer. The key design choice is that every quantity a learner is likely to vary is surfaced in a single configuration block rather than buried in the model code.

\begin{lstlisting}[language=Python, caption={Key hyperparameters in the DNN template}]
EPOCHS = 300
INPUT_SIZE = 1
HIDDEN_LAYERS = [64, 32]    # can be modified, e.g., [64, 32, 128]
OUTPUT_SIZE = 1
\end{lstlisting}

Here \texttt{HIDDEN\_LAYERS} sets the depth and width of the network, appending an entry such as \texttt{[64, 32, 128]} adds a layer. And \texttt{EPOCHS} sets the training length. The notebook then runs the full supervised-learning pipeline (data generation, splitting, training, evaluation, and visualization), so a learner studies the effect of these choices by editing one block and re-running rather than by reading a lengthy walkthrough.

\subsection{Function Approximation Results}

\begin{figure}[!htbp]
    \centering
    \subfloat[$f(x)=\sin(x)$]{\includegraphics[width=0.485\linewidth]{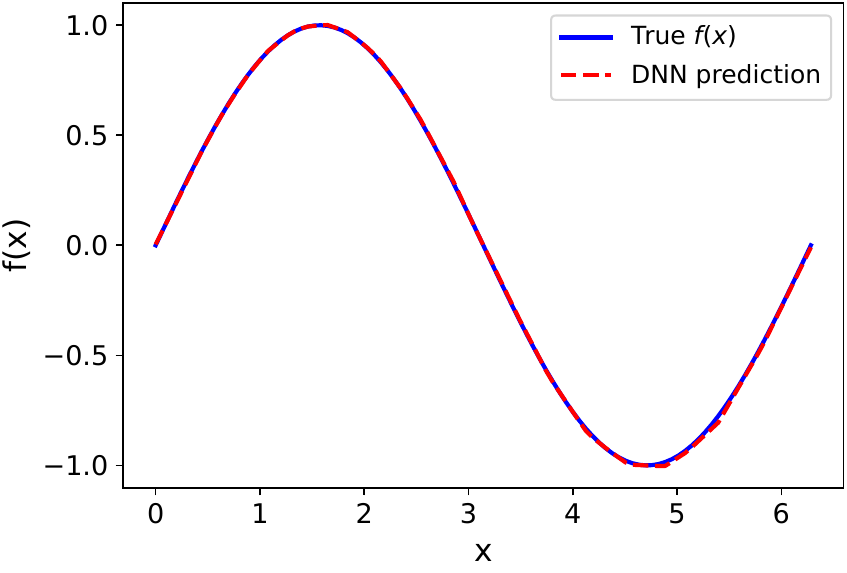}}\hfil
    \subfloat[$y=0.2x^3-0.5x^2+0.3x$]{\includegraphics[width=0.485\linewidth]{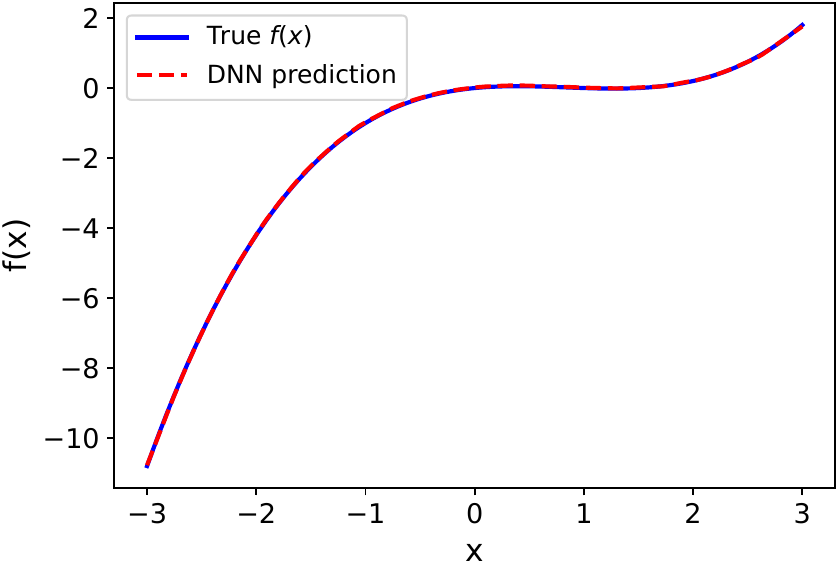}}
    \caption{Predicted versus true curves on unseen test data for two representative modes of the shared DNN template: (a) a trigonometric mapping and (b) a cubic polynomial. The identical network and training pipeline reproduce both.}
    \label{fig:tier1_fit}
\end{figure}

Evaluated on unseen test data, the same architecture and pipeline generalize across function families. Fig.~\ref{fig:tier1_fit} shows two representative modes: the predicted curve closely tracks the ground truth for both a trigonometric target and a cubic polynomial, confirming that the template is not tailored to any single mapping.

\subsection{Dataset and Load-Curve Fitting}
\label{subsec:loadcurve}

Keeping the same network and pipeline, the second mode (\texttt{DNN\_load\_curve}) replaces the analytic generator with \emph{sampled data}, in this case a measured load profile, so that the input--output relationship is learned from data rather than from a closed-form expression. Each feature--label pair $(x,y)$ is reshaped to the DNN interface, and the data are partitioned with the widely used $70/30$ train--test split, again exposed once in Section~0.

\begin{lstlisting}[language=Python, caption={Training configuration in Section~0 of \texttt{DNN\_load\_curve}}]
# === Section 0: Settings ===
TRAIN_RATIO = 0.70
EPOCHS = 500
HIDDEN_LAYERS = [64, 32, 128]   # can be changed to [64, 32] or other sizes
\end{lstlisting}

\begin{figure}[!htbp]
    \centering
    \includegraphics[width=0.75\linewidth]{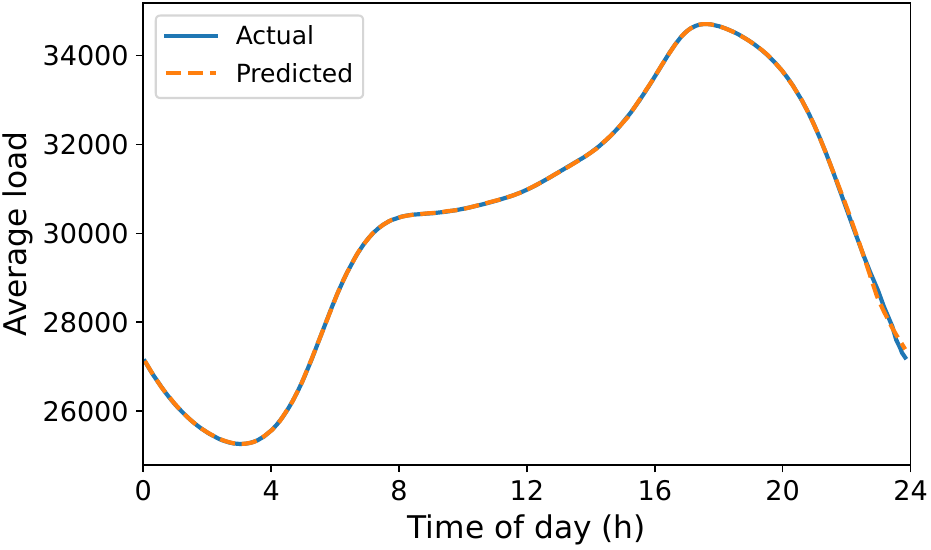}
    \caption{Load-curve learning with the \texttt{DNN\_load\_curve} module.}
    \label{fig:load_curve}
\end{figure}

Evaluating on the held-out $30\%$ reinforces the central lesson that predictive quality must be judged on unseen data. Fig.~\ref{fig:load_curve} shows the template reproducing a load curve from sampled data, confirming that the identical workflow transfers directly from analytic functions to measured signals. Overall, the foundational tier exposes the complete supervised-learning workflow through one configurable template, letting beginners experiment with architectures, hyperparameters, and data sources before advancing to the domain-coupled and frontier modules.

\section{Domain-Coupled Module (Tier~2): CNN Power-Flow Surrogate}
\label{sec:tier2}

The domain-coupled tier connects a neural network directly to a physics-based simulator. A CNN, implemented in \texttt{CNN\_for\_5\_bus\_system.ipynb}, is trained as a \emph{power-flow surrogate}: it maps known operating set-points (bus loads and generator dispatch) to the solved system state (bus voltage magnitudes and line power flows), so the nonlinear power-flow equations need not be solved online. The module shows how the Tier~1 workflow extends to structured, multi-output regression over quantities produced by a domain simulator; as before, the section emphasizes the design logic and code structure and leaves the hands-on detail to the notebook.

\subsection{Dataset Generation and Leakage-Free Design}

The modified PJM 5-bus system \cite{liSmallTestSystems2010} is loaded directly from \texttt{pandapower}. A single power-flow solution yields only one operating point, so the dataset is built by Monte-Carlo sampling around the nominal load: every load is perturbed by multiplicative Gaussian noise,
\begin{equation}
p^{(i)} = p^{\text{base}}\,\varepsilon, \qquad \varepsilon \sim \mathcal{N}(1,\sigma^{2}),
\end{equation}
with $\sigma = 10\%$ applied independently to the active and reactive components, and the flow is re-solved (non-converged points are discarded). A central design point, made explicit in the code, is the avoidance of data leakage. The inputs are restricted to the \emph{known} quantities: bus load ($P$, $Q$), generator active-power set-points, and generator voltage set-points, arranged as a structured $5\times4$ array. The solved voltages and line flows serve exclusively as targets.

\begin{lstlisting}[language=Python, caption={Dataset-generation settings in \texttt{CNN\_for\_5\_bus\_system}}]
N_SAMPLES = 100    # number of Monte-Carlo power-flow runs
LOAD_STD  = 0.10   # load noise std = 10% of the nominal demand
# inputs : load P, load Q, gen P, gen V-set -> (N, 5 buses, 4 features)
# targets: bus voltages (5), and line flows (P then Q)
\end{lstlisting}

The demo draws \texttt{N\_SAMPLES}$=100$ Monte-Carlo operating points and shuffles them into an $80/20$ train/test split; this lets learners see how dataset size, training length, and capacity jointly affect accuracy while keeping the notebook quick to run. Inputs are standardized with training-set statistics that are reused on the held-out test set.

\subsection{CNN Architecture, Model Variants, and Training}

The $5\times4$ feature map is convolved along the bus dimension, so the learned filters capture interactions among neighboring buses. The network stacks three one-dimensional convolutional (\texttt{Conv1D}) layers with rectified linear unit (ReLU) activations (kernel size two, valid padding), a flatten, a 12-unit dense layer, and a linear output; no pooling is used because the spatial extent (five buses) is small. The output width is either five (bus voltages) or $2\,n_{\text{line}}$ (per-line active and reactive flows). To let learners probe the effect of training length and width, four variants are defined (Table~\ref{tab:cnn_models}): Models~1--3 predict bus voltages and Model~4 predicts the line flows.

\begin{table}[!htbp]
\centering
\caption{CNN model variants used in the 5-bus surrogate module. Each variant uses three \texttt{Conv1D} layers; the filter counts are listed for the three layers in order.}
\label{tab:cnn_models}
\begin{tabular}{clcc}
\toprule
Model & Target & Epochs & Conv.\ filters \\
\midrule
1 & Bus voltage     & 50  & 24, 24, 12 \\
2 & Bus voltage     & 50  & 240, 240, 120 \\
3 & Bus voltage     & 200 & 240, 240, 120 \\
4 & Line power flow & 200 & 24, 24, 12 \\
\bottomrule
\end{tabular}
\end{table}

All variants train with an MSE loss and the Adam optimizer (batch size sixteen, $20\%$ validation split), tracking the mean absolute error (MAE) as an auxiliary metric. After training, the held-out test operating points, unseen during training and standardized with the training statistics, are predicted directly from the inputs, without invoking the power-flow solver, and compared against the reference solution.

\subsection{Results and Comparison}

\begin{figure}[!htbp]
    \centering
    \includegraphics[width=0.82\linewidth]{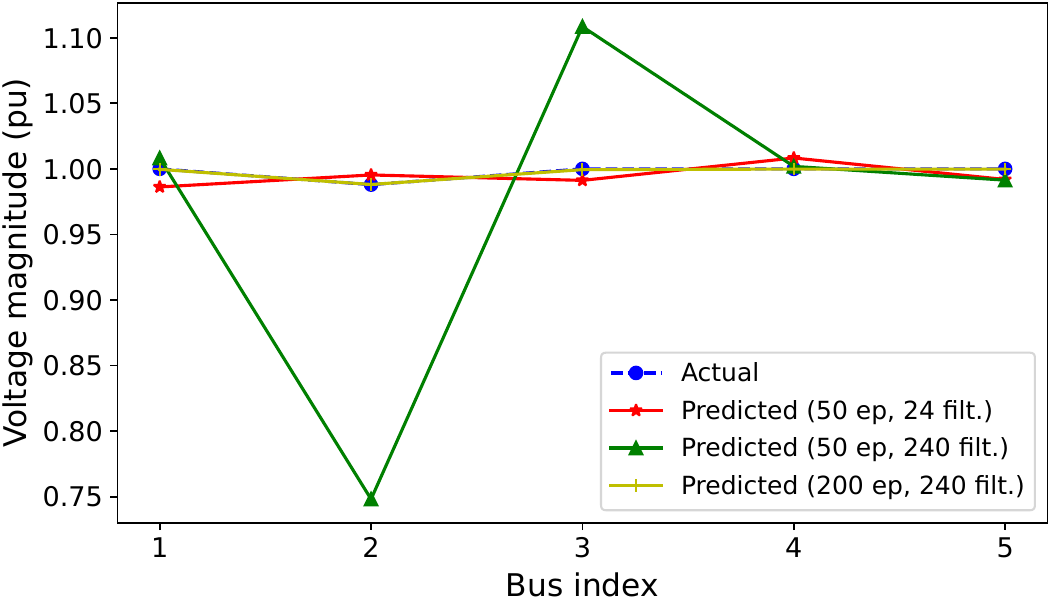}
    \caption{Actual versus predicted bus voltage magnitudes on an unseen test sample for the three voltage models (50 epochs/24 filters, 50 epochs/240 filters, and 200 epochs/240 filters).}
    \label{fig:cnn_voltage}
\end{figure}

The surrogate is evaluated on the held-out test operating points for bus voltages and for line active and reactive flows. A physical subtlety, made explicit in the notebook, shapes the voltage results: in \texttt{case5} the PV and slack buses are pinned to their voltage set-points, so only the single PQ-bus voltage actually varies across operating points, and the target is a near-flat $\approx 1.0$~pu profile with a small dip at that bus. For the voltage models (Fig.~\ref{fig:cnn_voltage}), the model trained for 200 epochs with 240 filters reproduces this profile almost exactly (test MAE below $10^{-3}$~pu), including the dip at the PQ bus, whereas the two lightly trained 50-epoch models incur more than an order-of-magnitude larger error: the 24-filter model wobbles around the flat reference, and the higher-capacity 240-filter model, undertrained after only 50 epochs, deviates most around the PQ bus. This gives an intuitive, hands-on illustration that model capacity must be paired with sufficient training length. Because the voltage target is nearly degenerate on this small system, the line flows below are the more demanding test of the surrogate.

\begin{figure}[!htbp]
    \centering
    \subfloat[Active power flow $P$]{\includegraphics[width=0.485\linewidth]{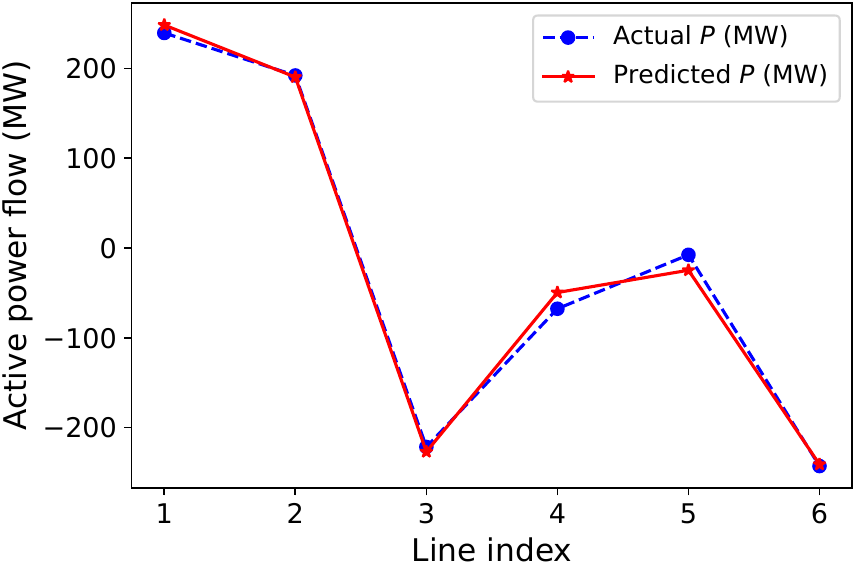}}\hfil
    \subfloat[Reactive power flow $Q$]{\includegraphics[width=0.485\linewidth]{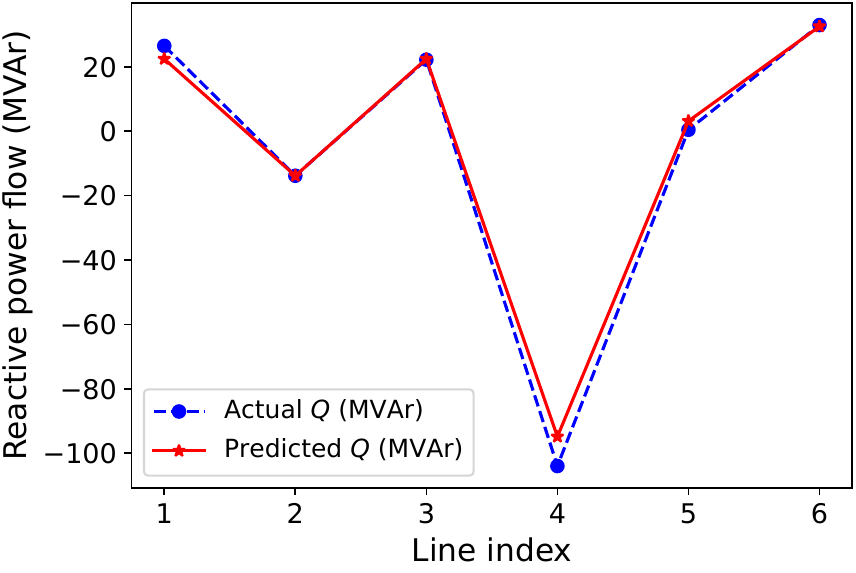}}
    \caption{Actual versus predicted line power flows on an unseen test sample: (a) active power $P$ and (b) reactive power $Q$.}
    \label{fig:cnn_flows}
\end{figure}

For the line flows (Fig.~\ref{fig:cnn_flows}), the power-flow model (Model~4) reproduces the active-flow pattern across all six lines, namely the large positive flows on the first two lines and the reversals on the remaining lines, with moderate per-line deviations (test MAE $\approx 14$~MW on flows spanning roughly $\pm 240$~MW). The reactive flows are likewise captured (test MAE $\approx 2$~MVAr), with the largest residual at the most heavily loaded line, whose sizable reactive excursion ($\approx -100$~MVAr) is slightly underestimated. This shows the surrogate learns the bulk power-flow solution from a small dataset while the largest, most nonlinear excursions remain the hardest to match, which is exactly the data-and-capacity trade-off these hands-on modules are meant to expose.

Overall, this module illustrates how the shared AI workflow extends to a domain-coupled surrogate-modeling task driven by physical simulations. Although the 5-bus system and the small demonstration dataset are intentionally lightweight, the same data-generation, modeling, and training principles scale directly to larger networks (e.g., the IEEE 39- or 118-bus systems) and to other engineering domains that rely on structured, simulation-generated data.

\section{Frontier Modules (Tier~3): Optimization, Control, and Physics}
\label{sec:tier3}

The frontier tier extends the AI application beyond supervised regression to three advanced method families: learned-constraint optimization, sequential decision making, and physics-informed learning.
Each subsection first identifies how AI learning is coupled to the engineering domain and highlights the corresponding executable modules, and then examines practical cases.

\begin{figure}[!htbp]
    \centering
    \includegraphics[width=\columnwidth]{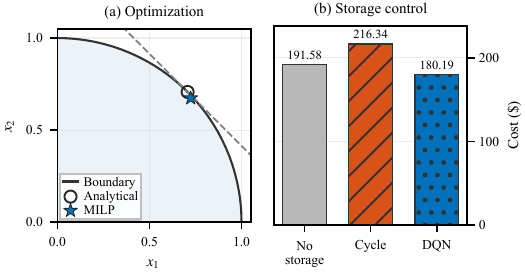}
    \caption{Executable example results for (a) DNN-assisted optimization and (b) DRL-based battery storage cost.}
    \label{fig:tier3_decision_results}
\end{figure}

\subsection{DNN-Assisted Optimization}
\label{subsec:dnnopt}

AI-assisted optimization places a learned mapping inside a mathematical program, allowing data-derived approximations of constraints that are difficult to express in a solver-compatible form.
This coupling extends the shared workflow from prediction to decision making.
In \texttt{DNN\_Assisted\_Opti.ipynb}, \texttt{ReLURegressor} approximates $g(\boldsymbol{x})=x_1^2+x_2^2-1$, \texttt{linear\_bounds} derives neuron preactivation limits over the decision box, and big-$M$ constraints embed the trained piecewise-linear mapping in a mixed-integer linear program (MILP).
Separate network and original-constraint evaluations distinguish exact encoding of the trained network from satisfaction of the nonlinear constraint.

The applilcation case maximizes $x_1+x_2$ subject to $g(\boldsymbol{x})\leq0$ over $[-1.5,1.5]^2$.
A 64-unit, one-hidden-layer ReLU network trained on 6000 samples.
As shown in Fig.~\ref{fig:tier3_decision_results}(a), the resulting MILP returns $\boldsymbol{x}^{\star}=(0.7237,0.6735)$, with objective 1.3973 and true constraint value $-0.02256$.
This example illustrates how nonlinear relationships in power system applications can be approximated by AI models and subsequently embedded as constraints in optimization-based decision models.

\subsection{Deep Reinforcement Learning for Battery Storage Control}
\label{subsec:drl}

DRL formulates battery scheduling as a finite-horizon Markov decision process, coupling neural function approximation to a domain environment through state transitions and rewards \cite{sheFusionMicrogridControl2023}.
\texttt{DRL\_BESS.ipynb} schedules a battery energy storage system (BESS) for cost minimization.
The normalized state comprises hour, time-of-use (ToU) price, load, and battery state of charge (SOC), while the actions are charging, idling, and discharging.
\texttt{BatteryStorageEnv.step} enforces power and energy limits, updates the SOC, and returns a scaled negative complete cost as the reward.
The \texttt{DQN} and \texttt{ReplayMemory} modules expose action-value approximation and experience replay, while the training block implements $\epsilon$-greedy exploration and target-network Bellman updates.
This separation preserves the model, training, and evaluation sequence while allowing the physical limits, reward terms, and learning settings to be modified independently.

For the default case, a deep $Q$-network (DQN) is trained for 300 episodes on one synthetic 24-h profile.
Fig.~\ref{fig:tier3_decision_results}(b) shows that its greedy training-day rollout has a complete cost of \$180.19, compared with \$191.58 without storage and \$216.34 for the straightforward cycle rule.
The 5.9\% reduction relative to the no-storage case is obtained while returning the 60-kWh battery to its initial empty state. 
Because every training episode and the reported rollout use the same synthetic profile and illustrative cost coefficients, this result verifies the executable state--action--reward coupling.
Beyond this example, the same formulation can be extended to sequential power system control tasks in which learned policies must account for system states, operating constraints, and customized objectives.

\subsection{Physics-Informed Neural Networks for the Swing Equation}
\label{subsec:pinn}

Physics-informed learning couples a neural approximation to known differential equations through the training objective, thereby extending the shared workflow from data fitting to power system dynamics \cite{raissiPhysicsinformedNeural2019,misyrisPhysicsInformedNeural2020}.
In \texttt{PINN\_Frequency.ipynb}, a two-output MLP maps time to rotor angle $\hat{\delta}(t)$ and angular speed $\hat{\omega}(t)$. The \texttt{physics\_residual} routine uses automatic differentiation to evaluate $r_{\delta}=\dot{\hat{\delta}}-\hat{\omega}$ and $r_{\omega}=M\dot{\hat{\omega}}+D\hat{\omega}+K_s\hat{\delta}-\Delta P$ at unlabeled collocation points.
A composite loss then combines dimensionless data, initial-condition, and residual terms, while the physical parameters, sample densities, and loss weights remain exposed for experimentation.

\begin{figure}[!htbp]
    \centering
    \includegraphics[width=\columnwidth]{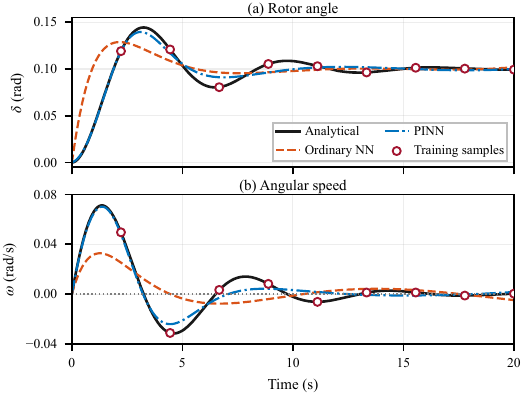}
    \caption{Dynamic trajectories from the analytical solution, ordinary neural network, and PINN.}
    \label{fig:tier3_pinn_trajectory}
\end{figure}

In the swing equation case, the ordinary neural network and PINN use identical architectures, initial weights, initial condition, and optimizer budgets.
Only the PINN adds 300 collocation points.
Fig.~\ref{fig:tier3_pinn_trajectory} compares both learned trajectories with the analytical response and the sparse training samples.
The PINN more closely reproduces the rotor angle swing and the angular speed excursion between the sparse samples.
For broader power system applications, embedding governing equations in the training objective can support sparse data modeling of dynamics relevant to stability analysis and to frequency-aware system operation~\cite{yinEncodingFrequencyDynamics2026, she2023virtual}.

\section{Discussion and Educational Impact}
\subsection{Addressing the Barriers Identified by the Survey}
The module library is designed as a direct response to the barriers reported in Section~\ref{sec:survey}. The survey found that a large majority of respondents encounter at least one obstacle before running an AI model and that generic, image-based examples are perceived as only weakly relevant to power engineering. The framework counters both issues. The unified template and one-click Colab execution remove environment- and setup-related barriers, so that a first result is obtained within minutes and without local installation. The explicit pairing of each AI concept with a power-system task replaces the abstract image-classification exemplar with problems that respondents recognize from their own work. The progressive tiering, in turn, lets learners enter at a level matched to their background and advance without switching frameworks.

\subsection{A Reusable Framework Beyond Power Systems}
Because every module is built on the same configurable template, the methodology transfers readily to other data-driven disciplines. The foundational regression workflow applies unchanged to problems in finance, weather, or biology; the surrogate-modeling pattern of Tier~2 applies to any expensive simulator; and the optimization, control, and physics-informed patterns of Tier~3 recur across engineering domains. Presenting these patterns through a shared interface promotes interdisciplinary learning and lets instructors reuse a single teaching infrastructure across courses.

\subsection{Integration with the Course, Webinar, and Community Feedback}
The modules complement the IEEE online course \emph{AI for Power and Energy Systems}~\cite{Onlinecourse2026} and the IEEE PES webinar series~\cite{liWebinarAIPower2026, liWebinarAIPowerslides2026}: the course and webinar provide the conceptual framing, while the notebooks provide the hands-on counterpart that learners run and modify. Google Colab support enables self-paced study on any machine, and the pinned environment specification makes the same notebooks suitable for graded assignments or laboratory sessions. Early deployment already yields encouraging feedback: with more than $590$ live attendees in Part~I, among the ten most-attended IEEE PES webinars. Also the open-source hands-on codes received  over $344$ repository visits within the first two weeks. The materials reach a large and engaged audience, echoing the survey-based demand and closing a course--webinar--survey--feedback loop that guides this work. This coupling of narrated concepts with executable, reproducible code is the mechanism by which the framework moves learners from passive consumption toward active experimentation.

\subsection{Toward Engineering-Grounded AI}
Beyond lowering the entry barrier, the framework points toward a broader design principle that we term \emph{engineering-grounded AI} (EGAI): AI systems for power and energy applications should be developed so that every stage of their operation (input, routine, skills, workflow, iteration, verification, and output) conforms to established engineering principles and power-system domain rules, rather than treating the model as a task-agnostic black box~\cite{yinAIAgentsTaskSimple2026}. Each module in the library embodies this view in miniature: inputs carry physical meaning (bus injections, states of charge, or system parameters), the workflow is constrained by domain structure (a power-flow solver, a Markov decision process, or the swing equation), and each output is explicitly verified against a physical or analytical reference before it is trusted. Making these engineering constraints visible and executable, rather than hiding them behind a conversational interface, is what turns a generic AI recipe into a dependable power-system tool, and it is the principle we encourage learners to carry from these examples into their own work.

\subsection{Limitations and Future Directions}
The present modules are intentionally lightweight, and their small demonstration settings are chosen for speed rather than realism; scaling them to larger systems and larger datasets is a natural next step already signposted in the notebooks. Further directions include adding time-series modules based on recurrent and long short-term memory networks for forecasting, incorporating explainability tools so that learners can interrogate model behavior, and extending the frontier tier toward digital-twin and closed-loop grid-operation workflows. A formal classroom evaluation of learning outcomes, following the survey-driven design reported here, would further quantify the educational impact.

\section{Conclusion}
This paper presented an educational framework and an open, executable module library for teaching AI in power systems. Grounded in a community survey of researchers and practitioners, the framework organizes six hands-on modules along a progressive difficulty ladder that maps core AI concepts onto representative power-system tasks. Across the three tiers, a single configurable notebook template carries the learner from foundational DNN function and load-curve fitting, through a domain-coupled CNN power-flow surrogate, to frontier modules on DNN-assisted optimization, DRL for storage control, and physics-informed learning of the swing equation.

The developed samples and codes lower setup barriers and anchor each method in a familiar power-system problem.
They are released as a reproducible, Colab-ready notebook delivered through an IEEE online course and an IEEE PES webinar with strong early engagement.
The proposed framework bridges conceptual understanding and practical implementation and encourages active experimentation over passive reliance on LLMs.

\section*{Acknowledgment}
The authors would like to thank the CURENT Engineering Research Center for facility support and the IEEE PES Artificial Intelligence for Power Systems Coordinating Committee (AIPSCC) for support in conducting the survey reported in this paper.
\ifCLASSOPTIONcaptionsoff
\newpage
\fi

\bibliographystyle{IEEEtran}
\bibliography{1} 
\end{document}